\documentstyle{elsart}
\catcode`\"=\active\let"=\"
\def\tree{{\sc Tree}}
\def\3{\ss }
\def\co{{\cal O}}
   \def\ApJ{Astrophys. J.}
   \def\ApJS{Astrophys. J. Suppl.}
   \def\AA{Astron. Astrophys.}
   
   \def\MN{Monthly Notices Roy. Astron. Soc.}
   \def\JC{Journ. Comp. Phys.}
   \def\AJ{Astron. J.}
   \def\Nat{Nature}
   \def\ASS{Astrophys. Space Sci.}

   \def\ARAA{Ann. Rev. Astron. Astrophys.}

   \def\PASJ{Publ. astr. Soc. Jap.}

   \def\ZA{Zeits. f. Astroph.}

   \def\NewA{New Astronomy}
   \def\CeMDA{Celest. Mech. Dyn. Astron.}
   \def\Ica{Icarus}
 \def\REF#1#2#3#4#5{#2, {\em #1}, {\bf #4} (#3) #5}

  \def\parder#1#2{{\partial #1\over\partial #2}}
 \def\dedet#1#2{\left(\delta #1\over \delta t\right)_{\rm #2}}
 \def\div#1{{1\over r^2}\parder{}{r}\left(r^2 #1 \right)}
 
 \def\s{\sigma}
 \def\sr{\sigma_r}
\begin{document}
\begin{frontmatter}
%
%next line: title of the article
\title{Direct $N$-body Simulations}
\footnote{to appear in: Riffert H., Werner K. (eds), Computational 
Astrophysics, The Journal of Computational and Applied Mathematics (JCAM),
Elsevier Press, Amsterdam.}
%
%next line: name of first author
\author[AUT1]{Rainer Spurzem}
%
%next line: address of first author
\address[AUT1]{Astronomisches Rechen-Institut, M"onchhofstra\3e 12-14,
D-69120 Heidelberg, Germany}

%next line: this is where the abstract goes
\begin{abstract}
Special high-accuracy direct force summation $N$-body algorithms and
their relevance for the simulation of the dynamical evolution of star
clusters and other gravitating $N$-body systems in astrophysics are
presented, explained and compared with other methods. Other methods means here
approximate physical models based on the Fokker-Planck equation
as well as other, approximate algorithms to compute the gravitational
potential in $N$-body systems. Questions regarding the parallel implementation
of direct ``brute force'' $N$-body codes are discussed. The astrophysical
application of the models to the theory of relaxing rotating and
non-rotating collisional star clusters is presented, briefly mentioning
the questions of the validity of the Fokker-Planck approximation, the
existence of gravothermal oscillations and of rotation and primordial
binaries.
\end{abstract}
\end{frontmatter}
 
%the main body of the text starts here
% ==> in the text you refer to this using
%      \cite{Boyce89}
%

\section{Introduction}
\label{sect2} 
 
``The dynamical evolution of an isolated spherical system composed of
very many mass points has an appealing simplicity. The Newtonian laws of
motion are exact, and all average quantities are functions only of radial
distance $r$ and time $t$. Nevertheless, it is only recently, with the
availability of fast computers, that a systematic understanding of how such
systems develop through time has emerged. Since these idealized systems
should provide a very good approximation for globular clusters in this
and other galaxies, the theory of their development is an important
part of astronomy as well as an interesting branch of theoretical
particle dynamics.'' \cite{Spitzer87}

Once celestial mechanics was one of the most important fields of 
astronomy. Nowadays astrophysics has become much wider in scope,
including fields like stellar astrophysics and gas and
plasma dynamics of interstellar matter. For some objects, however,
the pure dynamics of gravitating mass points still provides an excellent
description of the global dynamical evolution or gives at least
the dominating background in which the gaseous or baryonic matter
evolves. Such objects are, starting from the large scale, the
entire universe itself, some evolutionary phases of galaxies and
galactic nuclei, globular and open star clusters, and last but not
least our planetary system. Globular star
clusters are gas free systems with some $10^5$ stars, orbiting
around our own \cite{DjorgovskiM94} and other galaxies \cite{Milleretal97}. 

This article aims at the complex interplay of thermodynamic processes
like heat conduction and relaxation with the physics of self-gravitating
systems and the stochastic nature of star clusters having finite particle
number $N$, and the specific computational and physical models used
for the numerical simulation of the dynamical evolution of star clusters
under these processes on the computer.
Globular clusters are a very good laboratory for relaxation processes in
discrete particle systems,
because their dynamical and relaxation
timescales are well separated from each other and from the lifetime
of the cluster and of the universe as a whole. In this article the
methods appropriate to
model their evolution are in the focus.
Other kinds of $N$-body simulations are useful
for example for hydrodynamics (``smoothed particle hydrodynamics''), 
galaxy dynamics (``collisionless systems'') or cosmological
$N$-body simulations of structure formation in the universe and are
covered by other articles in this volume. The main distinction of those
from the models presented here, is that the dynamics of systems dominated
by two-body relaxation (``collisional systems'') requires typically
very high accuracies (typical energy error per crossing time 
$\Delta E/E < 10^{-5}$ or smaller) over very long physical integration
times (thousands of crossing times). The term ``collisional'' here
always refers to systems, whose evolution is influenced by relaxation
through elastic two- or more-body encounters, {\it not} to physical
collisions, where two stars collide and merge or disrupt each other.
As a consequence of the high accuracy requirements for
collisional $N$-body simulations, commonly known
algorithms like the leap-frog time integration and the \tree-method to
compute the gravitational potential of a particle distribution, are
not efficient to use here; the use of high-order time integration schemes
and ``brute--force'' algorithms to compute the potential are more
efficient, as will be argued below.

This article is organized as follows: this introduction is followed
by a section on the approximate models of self-gravitating collisional
$N$-body systems, after which practical and theoretical aspects of
the corresponding highly accurate direct $N$-body simulations are
presented. Finally astrophysical applications of the methods and
relevant questions under study are presented.

Let us begin with the definition of some useful time scales.
A typical particle
crossing time $t_{\rm cr}$ in a star cluster is

\begin{equation}
t_{\rm cr} = {r_h \over \sigma_h}\ , 
\label{1.1} 
\end{equation}

where $r_h$ is the radius containing 50 \% of the total mass and
$\sigma_h$ is a typical velocity associated
with the root mean square random motion (velocity dispersion) taken
at $r_h$. If virial equilibrium prevails, we have
$\sigma_h^2 \approx GM_h/r_h$ (where the sign $\approx$ here and
henceforth means ``approximately equal'' or ``equal within an order
of magnitude''), thus

\begin{equation}t_{\rm cr} \approx \sqrt{r_h^3\over G M_h} \ . 
\label{1.2} 
\end{equation}

This is equal to the dynamical timescale, which is also used for example
in the theory of stellar structure and evolution. Global dynamical
adjustments of the system, like oscillations, are connected
with this timescale. Taking the square of equation \ref{1.2} yields
$t_{\rm cr}^2 \approx r_h^3/(GM_h)$ which is related to Kepler's third law,
because the orbital velocity in a Keplerian point mass potential has the
same order of magnitude as the velocity dispersion in virial equilibrium.

Unlike most laboratory gases stellar systems are not usually in
thermodynamic equilibrium, neither locally nor globally. 
Radii of stars are usually extremely small
relative to the average interparticle distances of stellar
systems (e.g. the radius of the
sun is $r_\odot\approx 10^{10}$ cm, a typical distance between
stars in our galactic neighbourhood is of the order of $10^{18} $cm).
Only under rather special conditions in the centres of galactic nuclei
and during the short high-density core collapse phase of a 
globular cluster, stellar densities might become large enough that
stars come close enough to each other to collide, merge or disrupt
each other.

Therefore it is extremely unlikely under normal conditions that two stars
touch each other during an encounter; encounters
or collisions usually are elastic gravitative scatterings. 
Fairly generally the mean interparticle distance is large
compared to $p_0=Gm/\s^2$, which is the impact parameter for a 
$90^o$ deflection in a typical encounter of two stars
of equal mass $m$, where the relative velocity
at infinity is $\sqrt{2}\s$,
with local 1D velocity
dispersion $\s$. Thus most encounters are small-angle deflections.
The relaxation time $t_{\rm rx}$ 
is defined
as the time after which the root mean square velocity increment due
to such small angle gravitative deflections is of the same order as
the initial velocity dispersion of the system. We use
the local relaxation time as defined by \cite{Larson70}:

\begin{equation}
t_{\rm rx} = {9\over 16 \sqrt{\pi}} {\sigma^3 \over G^2 m \rho 
    \ln(\gamma N)} \ . 
\label{1.3}
\end{equation}

$G$ is the gravitational constant, 
$\rho$ the mean stellar mass density, $N$ the total particle number,
and $\gamma$ a parameter of order unity, which results from an
integration over all possible impact parameters for a two-body
encounter. Taking the linear system dimension as a maximum 
impact parameter
yields $\gamma = 0.4$ \cite{Spitzer87}. 
Measurements in direct star by star evolutionary simulations of
stellar systems are more in favour of $\gamma = 0.11 $ \cite{GH1,GH2},
which is the value used throughout this article. 

Assuming virial equilibrium a fundamental proportionality turns out:

\begin{equation}
{t_{\rm rx}\over t_{\rm dyn}} \propto  {N\over \ln(\gamma N)} \ .
\label{1.4} 
\end{equation}

(cf. e.g. \cite{Spitzer87}). As a result, for very large $N$, dynamical
equilibrium is attained much faster than thermodynamic equilibrium.
If one assumes a purely kinetic temperature definition, it ensues that
in star clusters the temperatures (or velocity dispersions) can remain
different
for different coordinate directions over many
dynamical times. For example, in a spherical
system the radial and tangential velocity dispersion would be
different, which is denoted as anisotropy. 

There are several reasons to believe that anisotropy is present and
important for the dynamical evolution of astrophysical star clusters.
Many
observations are matched better by models including anisotropy
\cite{LuptonGG87},
and all direct simulations exhibit the formation
of anisotropy under very general conditions, independent of
the underlying physical cause driving the system's evolution. 
\cite{BettwieserSp86} showed in the context
of a gas dynamical model of star clusters, that isotropy 
remains only under very special conditions (linear profiles
of velocities of mass and energy transport), and a similar study 
\cite{HenslerSpBT95} gives the same result for axisymmetric
collapsing gaseous systems.

\section{Approximate Models}

\subsection{Fokker-Planck Approximation}

Unfortunately,
the direct simulation of such rich stellar systems as globular clusters
with star-by-star modelling is not yet possible. The gap between the
largest useful $N$-body models with particle numbers of the
order of a few $10^4$ particles
and the median globular
star cluster ($N \sim 5\times 10^5$) can only be bridged at present by
use of theory.  There are two main classes of theory: (i) Fokker-Planck
models, which are based on the Boltzmann equation of the kinetic theory
of gases \cite{CohnHW89,MurphyCH90,Grabhornetal92,Drukier95}, 
and (ii) gas models \cite{SpA,SpB,SpH},
which can be thought of as a set of
moment equations of the Fokker-Planck model.

These simplified models are the only detailed models which are directly
applicable to large systems such as globular clusters. But their
simplicity stems from many approximations and assumptions which are
required in their formulation.  Examples are the assumptions of
spherical symmetry, which
contradicts the asymmetry of the galactic tidal field, 
or statistical estimates of
cross sections for the formation of close binaries by three-body
or dissipative (tidal) two-body encounters, and for their subsequent
gravitational interactions with field stars. Such processes play
a dominant role to reverse core collapse of globular clusters,
which otherwise would inevitably lead to a singular density profile
with infinite density at the centre
\cite{BettwieserSu84,ElsonHI87,Hut93}.
 
The Fokker-Planck approximation truncates the so--called B${}^2$GKY hierarchy
of kinetic equations (see \cite{BinneyT87}) at lowest order assuming
that for most of the time all particles are uncorrelated with each
other and only coupled via the smooth global gravitational potential.
Correlations only play a role as a sequence of uncorrelated two-body
encounters. Instead of determining a general correlation function one
resorts to a phenomenological description of the effects of collisions
by computing diffusion coefficients directly from the known solution
of the two-body problems. Diffusion coefficients $D(\Delta v_i)$ and
$D(\Delta v_iv_j)$ denote the average rate of change of $v_i$ and
$v_iv_j$ due to the cumulative effect of many small angle deflections
during two-body encounters. Let $m$, $\vec{v}$ and $m_f$, $\vec{v_f}$
be the mass and velocity of a star from a test and field star
distribution, respectively (both distributions can but 
need not to be the same).
In Cartesian geometry the diffusion coefficients are defined by

\begin{equation}
 D(\Delta v_i)  = 4\pi G^2 m_f \ln\Lambda \parder{h(\vec{v})}{v_i}
 \ \ ;\ \ D(\Delta v_iv_j)  = 4\pi G^2 m_f \ln\Lambda 
    {\partial^2 g(\vec{v}) \over \partial v_i v_j} \ \ ;
\label{2.1.20} 
\end{equation}

where $f$ is the phase space density of stars (briefly: distribution
function) and $g$, $h$ are
the Rosenbluth potentials defined in \cite{RMJ57}

\begin{equation}
 h(\vec{v})  = (m+m_f)
 \int {f(\vec{v}_f)\over\vert\vec{v}-\vec{v}_f\vert}
 d^3\!\vec{v}_f \ \ ;\ \
 g(\vec{v})  = m_f \int f(\vec{v}_f) \vert\vec{v}-\vec{v}_f\vert
 d^3\!\vec{v}_f \ . 
\label{2.1.21} 
\end{equation}

Note that provided the distribution function $f$ is given in terms
of a convenient polynomial series as in Legendre polynomials the
Rosenbluth potentials can be evaluated analytically to arbitrary
order, as was seen already by \cite{RMJ57}, see for a modern rederivation
and its use for star cluster dynamics \cite{SpD,SpF}.
With these results we can finally write down the local Fokker-Planck
equation in its standard form 
for the Cartesian coordinate system of the $v_i$:

\begin{eqnarray}
\label{2.1.22}
\parder{f}{t} &+& \vec{v}_i\parder{f}{\vec{r}_i}+\vec{{\dot v}}_i
\parder{f}{\vec{v}_i} = \dedet{f}{enc} \ ; \\
\dedet{f}{enc} &=&
   - \sum_{i=1}^3 \parder{}{v_i}\Bigl[
        f(\vec{v}) D(\Delta v_i) \Bigr]
    + {1\over 2} \sum_{i,j=1}^3
       {\partial^2\over\partial v_i\partial v_j}\Bigl[
        f(\vec{v})
         D(\Delta v_iv_j)
       \Bigr] \ . 
\label{2.1.23}
\end{eqnarray}

The subscript ``enc'' should refer to encounters, which are the driving
force of two--body relaxation.
Still Eq. \ref{2.1.22} is a six-dimensional integro-differential equation; its 
direct numerical simulation in stellar dynamics can presently only be
done by further simplification. First Jeans' theorem is applied and
$f$ transformed into a function of the classical integrals of motion of
a particle in a potential under the given symmetry, as e.g. energy $E$
and modulus of the angular momentum $J^2$ in a spherical potential or
$E$ and $z$-component of angular momentum $J_z$ in axisymmetric coordinates.
Thereafter the Fokker-Planck equation can be integrated over the accessible
coordinate space for any given combination of constants of motion and
the orbit-averaged Fokker-Planck equation ensues. By transformation from
$v_i$ to $E$ and $J$ and via the limits of the orbital integral the
potential enters both implicitly and explicitly. In a two-step scheme
alternatively solving the Poisson- and Fokker-Planck equation a
direct numerical solution is obtained 
\cite{Cohn80,Drukier95,Takahashi95,Takahashi96,Takahashi97,EinselSp99}.
One of the main uncertainties in this method is
that for non-spherical mass distributions the orbit structure in the
system may depend on unknown non-classical third integrals of motion
which are neglected.

\subsection{Anisotropic Gaseous Model}
\label{sect2.3}

The local Fokker-Planck equation Eq. \ref{2.1.22} is utilized in another
way for gaseous or conducting sphere models of star clusters. Integrating
it over velocity space with varying powers of the velocity coordinates
yields a system of equations in the spatial coordinates; the local
approximation is used in the sense that the orbit structure of the
system is not taken into account, diffusion coefficients and all
other quantities are assumed to be well defined just as a function of
the local quantities (density, velocity dispersions and so on). The system
of moment equations is truncated in third order by a phenomenological equation
of heat transfer. Such approach has been suggested by 
\cite{LyndenBE80,Heggie84} and generalized to anisotropic systems
by \cite{BettwieserSp86,SpA}, and for a presentation of the recent
model see e.g. \cite{SpD}. In the following the derivation of the model
equations is decribed.

\subsubsection{The ``Left Hand Sides''}
\label{sect2.3.1}

In spherical symmetry, polar coordinates $r$
$\theta $, $\phi$
are used and $t$ denotes the time. The vector
$\vec{v} = (v_i), i=r,\theta,\phi$, denotes the velocity
in a local Cartesian coordinate system at the spatial point
$r,\theta,\phi$. For brevity
$u=v_r$, $v=v_\theta$, $w=v_\phi$ is used. The distribution function $f$,
which due to spherical symmetry is a function of $r$, $t$,
$u$, $v^2+w^2$ only, is normalized according to

\begin{equation} 
\rho(r,t) = \int f(r,u,v^2+w^2,t) du\,dv\,dw,   
\label{2.2.1} 
\end{equation}

where $\rho(r,t)$ is the mass density; if $m$ denotes
the stellar mass, we get the particle density $n=\rho/m$. Then

\begin{equation} 
{\bar u} = \int u f(r,u,v^2+w^2,t) du\,dv\,dw,   
\label{2.2.2} 
\end{equation}

is the bulk radial velocity of the stars. 
Note that for the analogously defined quantities ${\bar v}$ and
${\bar w}$ we have
${\bar v} = {\bar w} = 0$.

 In order to go ahead to the anisotropic gaseous model equations
 we now turn back to the left hand side of the Fokker-Planck
 equation Eq. \ref{2.1.22}, 
 which is the collisionless Boltzmann or Vlasov operator.
 For practical reasons we prefer for the left hand side
 local Cartesian velocity coordinates, whose
 axes are oriented towards the $r$, $\theta$, $\phi$ coordinate space
 directions.
 With the Lagrange function 

 \begin{equation}
{\cal L} = {1\over 2}\bigl({\dot r}^2 + r^2{\dot\theta}^2 +
         r^2 \sin^2\!\!\theta\, {\dot\phi}^2\bigr) - \Phi(r,t) 
\label{2.3.1} 
\end{equation}

 the Euler-Lagrange equations of motion for a star moving in
 the cluster potential $\Phi$ become:

 \begin{equation}
  {\dot u}  = - \parder{\Phi}{r} + {v^2\!+\!w^2\over r} \ \ ;\ \
  {\dot v}  = - {uv\over r} + {w^2\over r\tan\theta} \ \ ;\ \
  {\dot w}  = - {uw\over r} - {vw\over r\tan\theta} \ \ .
 \label{2.3.2} 
\end{equation}

 The complete local Fokker-Planck equation, derived from Eq. \ref{2.1.22},
 attains the form

\begin{equation}
\parder{f}{t} + u\parder{f}{r} + {\dot u}\parder{f}{u} +
     {\dot v}\parder{f}{v} + {\dot w}\parder{f}{w} = \dedet{f}{enc}\ ,
\label{2.3.3} 
\end{equation}

 where the term subscribed by ``enc'' denotes the terms
 involving diffusion coefficients as in Eq. \ref{2.1.23}.
 Moments $\langle i,j,k \rangle $ of $f$ are defined in the
 following way (all integrations range from $-\infty $ to $\infty $):

 \begin{eqnarray}
 \langle 0,0,0\rangle &:=& \rho = \int f dudvdw  \ \ ; \ \
 \langle 1,0,0\rangle  :=  \bar{u} = \int uf dudvdw \\
 \langle 2,0,0\rangle &:=& p_r + \rho\bar{u}^2 = \int u^2 f dudvdw \\
 \langle 0,2,0\rangle &:=& p_\theta = \int v^2 f dudvdw \ \ ; \ \
 \langle 0,0,2\rangle  :=  p_\phi = \int w^2 f dudvdw \\
 \langle 3,0,0\rangle &:=&  F_r + 3\bar{u}p_r + \bar{u}^3 = 
                            \int u^3 f dudvdw \\
 \langle 1,2,0\rangle &:=&  F_\theta + \bar{u}p_\theta = \int uv^2 f dudvdw \\
 \langle 1,0,2\rangle &:=&  F_\phi + \bar{u}p_\phi = \int uw^2 f dudvdw\ . 
 \label{2.3.4} 
\end{eqnarray}

 Note that the definitions of $p_i$ and $F_i$ are such that they
 are proportional to the random motion of the stars. Due to spherical
 symmetry we have $p_\theta = p_\phi =: p_t$ and 
 $F_\theta = F_\phi =: F_t/2$. 
 By $p_r = \rho\sr^2$ and $p_t = \rho\s_t^2$
 the random velocity dispersions are given, which are closely related
 to observables in globular star clusters and galaxies. 
 It is convenient to define velocities of energy transport
 by 
 \begin{equation}
 v_r  = {F_r \over 3 p_r} + u \ \ ; \ \
 v_t  = {F_t \over 2 p_t} + u \ .
\label{2.3.5} 
\end{equation}
 By multiplication of the Fokker-Planck equation \ref{2.3.3} with
 various powers of $u$, $v$, $w$ we get up to second order the
 following set of moment equations (for a detailed derivation
 in the here used variables see \cite{Spurzem88}, bar for
 $\bar{u}$ dropped in the following):
 \begin{eqnarray}
 \parder{\rho}{t} + \div{u\rho} &=& 0 \\
 \label{2.3.6a}
 \parder{u}{t}+u\parder{u}{r} + {GM_r\over r^2} +
     {1\over\rho}\parder{p_r}{r} + 2{p_r - p_t\over\rho r} &=& 0 \\
 \label{2.3.6b}
 \parder{p_r}{t} + \div{u p_r} +
 2 p_r \parder{u}{r} + \div{F_r} - {2F_t\over r} &=&
  \dedet{p_r}{enc,bin3} \\
 \label{2.3.6c}
  \parder{p_t}{t} + \div{u p_t} +
 2 {p_t u\over r} + {1\over 2}\div{F_t} + {F_t\over r} &=& 
  \dedet{p_t}{enc,bin3} \ .
 \label{2.3.6d} 
\end{eqnarray}
 The terms labeled with ``enc'' and ``bin3'' symbolically denote
 the collisional terms resulting from the moments of the right
 hand side of the Fokker-Planck equation (Eq. \ref{2.1.23}) 
 and an energy generation
 by formation and hardening of three body encounters. Both will
 be discussed below. 
 With the definition of the mass $M_r$ contained in a sphere
 of radius $r$
 \begin{equation} 
\parder{M_r}{r} = 4 \pi r^2 \rho  
\label{2.3.7} 
\end{equation}
 the set of Eqs. \ref{2.3.6a}--\ref{2.3.6d} 
 is equivalent to gasdynamical equations
 coupled with Poisson's equation.
 Since moment equations of order $n$ contain moments of order
 $n\!+\!1$, it is necessary to close the system of the above equations
 by an independent closure relation. Here we choose the heat
 conduction closure, which consists of a phenomenological
 ansatz in analogy to gas dynamics. It was first used (restricted
 to isotropy) by \cite{LyndenBE80}. It is assumed
 that heat transport is proportional to the temperature gradient,
 where we use for the
 temperature gradient an average velocity 
 dispersion $\sigma^2 = (\sr^2 + 2\s_t^2)/3$
 and assume $v_r = v_t$ (this latter closure was first introduced
 by \cite{BettwieserSp86}). Therefore, the last two equations
 to close our model are

 \begin{equation}
  v_r - u + {\lambda\over 4\pi G\rho t_{\rm rx}} \parder{\s^2}{r} = 0 
  \ \ ; \ \ v_r = v_t  \ .
\label{2.3.11} 
\end{equation}

 With Eqs. \ref{2.3.6a}--\ref{2.3.6d}, \ref{2.3.7}, and \ref{2.3.11} we
 have now seven equations for our seven dependent variables
 $M_r$, $\rho$, $u$, $p_r$, $p_t$, $v_r$, $v_t$. 

\subsubsection{Binary Heating}
\label{sect2.3.2}

 It was already early realized that in a star cluster with
 single stars under high density conditions, one or more strongly
 bound binaries form, which could dominate the further evolution
 \cite{Henon71,AarsethHW74}. This is a contradiction
 to the basic assumption underlying the Fokker-Planck equation, that
 the only correlations in the system are those produced by a sequence
 of uncorrelated small-angle gravitational encounters. Nevertheless
 \cite{BettwieserSu84} introduced
 a phenomenological heat source into their gaseous model equations,
 in order to describe the input of random kinetic energy (``heat'')
 to the cluster by formation and hardening of so-called three-body
 binaries. 
 The ansatz for the functional form of the heating
 term has been clarified and more thoroughly discussed by
 \cite{Goodman87,HeggieR89}. They describe a simple
 estimate for the rate of formation of binaries by close three-body
 encounters of single stars; in subsequent superelastic scatterings
 between the formed binary and field stars the binary will on
 average become harder, provided its binding energy is large compared
 to the mean temperature of the surrounding single stars. Surplus
 kinetic energy taken from the gravitational binding energy of the
 binary members goes to the field star and thus provides a heating
 source for the core of the cluster. There
 is an upper limit of the binary binding energy given by the
 condition that the recoil on the binary in a typical
 superelastic scattering due to conservation
 of linear momentum in the process leads to escape of the binary.
 As a result each binary after its formation supplies a certain
 amount of energy by three-body encounters to the system until it
 escpes. The resulting heating term is
 (isotropic binary heating assumed):

 \begin{equation}
 \dedet{p_r}{bin3} = {2\over 3} C_b mn^3 \s^3 \Bigl({Gm\over\s^2}\Bigr)^5 
 \ ;\ \ \ \ \ \dedet{p_t}{bin3} = \dedet{p_r}{bin3} 
 \label{2.3.16} 
 \end{equation}

 Here a simple estimate using gravitational focusing and the probability
 that three particles come together have been employed.
 $C_b$ is a constant of proportionality
 which ies expected to have a value between 75 and 90 for
 an equal mass system; for more details
 see the above cited papers.

\subsubsection{The ``Right Hand Sides''}
\label{sect2.3.3}

All right hand sides of the moment equations \ref{2.3.6a}--\ref{2.3.6d} are
calculated by multiplying the right hand side
(the encounter term) of the Fokker-Planck equation as it occurs
in Eq. \ref{2.1.23} with the appropriate powers of $u$, $v$ and $w$ and
integrating over velocity space. 
There is only one non-trivial encounter term to be determined
for the collisional decay of anisotropy. It is self-consistently
computed by assuming
a certain Legendre series evaluation for $f$ up to second order
(i.e. including anisotropy) in the Appendix of \cite{SpD},
the result being ($p_a = p_r - p_t$):

\begin{equation}
\dedet{p_a}{enc} = -{p_a \over t_a} \ \ ;\ \
      t_a = {10\over 9} t_{\rm rx} \ \ ;\ \
t_{\rm rx} = {9\over 16\sqrt{\pi}} {\s^3\over G^2m\rho\ln(\gamma N)}
\ .
\label{2.3.18} 
\end{equation}

$t_a$ defined in the above equation denotes the characteristic decay
time of anisotropy; $t_{\rm rx}$ is equivalent to the standard two--body
relaxation time. The particular factors applied to it originate
unambigously from the Fokker-Planck collisional term evaluation with
the assumption of a certain normalization and functional form of $f$ by
a Legendre series. The procedure can be thoroughly followed in 
\cite{Larson70}.
For the above result terms quadratic in $p_a$ have been omitted. 
Comparisons with direct $N$-body
simulations suggest a more general ansatz 

\begin{equation}
\dedet{p_a}{enc} = -{p_a \over\lambda_a t_a} 
\label{2.3.20} 
\end{equation}

and it is shown that $\lambda_a = 0.1$ provides the best results 
\cite{SpD}. Sect. 4 describes some examples how well the gaseous and
Fokker-Planck models describe a star cluster's evolution as compared
to a direct $N$-body simulation. There is no other way to check the
theoretical models on the Fokker-Planck equation, because
the timescale for exponential
instability and deterministic chaos to occur in a self-gravitating
star cluster consisting of many stars of equal or at least similar
mass is of the same order as
a crossing time \cite{GoodmanHH93}. There is no analytical or
semianalytical
general solution of the $N$-body problem available 
in that case for the unperturbed problem. In contrast to this
in the case of solar system studies there is a semianalytic secular theory
\cite{Laskar89}) to be compared with the direct orbit integrations (see
e.g. \cite{LaskarQT92}). Here, for the star cluster case,
we only can rely on the comparison
of the numerical solutions obtained from different physical models,
as there are direct $N$-body integrations and models based on the
Fokker-Planck approximation.

\section{Direct $N$-body Simulations -- Methods and Algorithms}
\label{sect4}

\subsection{Introduction -- Density and Potential Computation}
\label{sect4.1}

To integrate the orbits of particles in time under their mutual
gravitational interaction the total gravitational potential at each
particle's position is required. Poisson's equation in integral
form gives the potential $\Phi$ generated at a point in coordinate
space $\vec{r}$ due to a smooth mass distribution $\rho(\vec{r})$
\begin{equation}
\Phi(\vec{r}) = - G \int {\rho(\vec{r}^\prime)\over
      \vert\vec{r}^\prime - \vec{r}\vert} d^3\vec{r}^\prime  \ \ .
\label{4.1}
\end{equation}
There are two fundamentally different methods to define the density
distribution as a function of a given particle distribution. The first
is based on a mesh in coordinate space; particles
are sampled on the mesh and their mass divided by the cell volume, which
provides a local density. This method called particle-mesh
requires for good statistics a sufficient number of
particles in each cell. There is
no or very little intrinsic particle-particle relaxation with this
method, but
there is relaxation of particle energies due to the finite resolution
of the mesh (see \cite{HockneyE88}, and for a more recent cosmological
application compare \cite{KlypinH97} and references therein). Refinements,
by which particles are smeared out by low-order interpolation formulae
(e.g. cloud-in-cell, CIC) or the acceleration
is interpolated within the cells
(e.g. Superbox \cite{Fellhaueretal98})
are possible and reduce mesh relaxation.

The second method is based on the particles itself. A kernel function
$W(\vec{x},h)$ is defined, normalized by $\int W d^3\vec{x} = 1$, where
$h$ describes a typical length scale over which the influence
of a particle decays. Therewith a
sampled density $\rho_s$ is defined in a mesh-free way as

\begin{equation}
\rho_s(\vec{r}) = \int W(\vec{r} -\vec{r}^\prime,h) 
\rho(\vec{r}^\prime)d^3\vec{r}^\prime
\end{equation}

where $h$ is a characteristic smoothing length.
A discrete particle distribution is given by
$\rho(\vec{r}) = \sum\delta(\vec{r}-\vec{r}_j) $ with $N$
particles distributed at positions $\vec{r}_j$. Hence we get a sampled
density from the discrete particle distribution as

\begin{equation}
\rho_s(\vec{r}) = \sum_{j=1}^N m_j W(\vec{r}-\vec{r}^\prime,h) 
\end{equation}

As an estimate for the density and other thermodynamic quantities
this method is used by ``smoothed particle
hydrodynamics'' simulations \cite{Monaghan92}, using kernel functions
$W$ with compact support, which means they are non-zero only in a
bounded volume. Here,
it is not intended to explain this further, the reader is referred
to the literature and other papers of this volume. If one takes
a $\delta$-function for the kernel as well and puts this into the
integral Poisson equation (\ref{4.1}) Newton's law turns out again:

\begin{equation}
\Phi(\vec{r}) = - G \sum_{j=1}^N {m_j\over \vert\vec{r}-\vec{r}_j\vert}
\end{equation}

Certainly this could have been written down immediately, but the above
description makes it easier to understand the relation of the direct
potential summation to the other methods. 
In the following, the prototype $N$-body integration method
using the above ``brute-force'' method, in a specific form called the
Hermite scheme \cite{MakinoA92}, 
shall be described in some more detail.
It is the most commonly used method in the field of globular cluster
dynamics and other studies requiring very high accuracies. 

\subsection{The Hermite Scheme}

Assume a set of $N$ particles with positions $\vec{r}_i(t_0)$ and
velocities $\vec{v}_i(t_0)$ ($i=1,\ldots, N$)
is given at time $t=t_0$, and let us look at a selected
test particle at $\vec{r} = \vec{r}_0=\vec{r}(t_0)$ and 
$\vec{v} = \vec{v}_0 = \vec{v}(t_0)$. Note that here and in the following
the index $i$ for the
test particle $i$ and also occasionally
the index $0$ indicating the time $t_0$
will be dropped for brevity; sums over $j$
are to be understood to include all $j$ with $j\ne i$, since there
should be no self-interaction.
Accelerations $\vec{a}_0$
and their time derivatives ${\bf{\dot a}}_0$ are calculated
explicitly:
\begin{equation}
 \vec{a}_0 = \sum_j Gm_j {\vec{R}_j\over R_j^3} \ \ ; \ \
 \vec{{\dot a}}_0 = \sum_j Gm_j \Biggl[
  {\vec{V}_j\over R_j^3} - {3 (\vec{V}_j\cdot \vec{R}_j)
   \vec{R}_j \over R_j^5} \Biggr] \ , 
\label{4.1.1} 
\end{equation}
where $\vec{R}_j:=\vec{r} - \vec{r}_j$, 
$\vec{V}_j := \vec{v} - \vec{v}_j$, $R_j:=\vert\vec{R}_j\vert$,
$V_j:=\vert\vec{V}_j\vert $.
By low order predictions,
\begin{eqnarray}
 \vec{x}_p(t) &=& {1\over 6}(t-t_0)^3\vec{{\dot a}}_0
          +{1\over 2}(t-t_0)^2\vec{a}_0 + (t-t_0)\vec{v} + \vec{x} \ ,\\
 \vec{v}_p(t) &=& {1\over 2}(t-t_0)^2\vec{{\dot a}}_0
          + (t-t_0)\vec{a}_0 + \vec{v} \ ,
\label{4.1.2} 
\end{eqnarray}
new positions and velocities for all particles at $t>t_0$ are calculated
and used
to determine a new acceleration and its derivative directly
according to Eq. \ref{4.1.1} at $t=t_1$, denoted by
$\vec{a}_1$ and $\vec{{\dot a}}_1$.
On the other hand $\vec{a}_1$ and $\vec{{\dot a}}_1$ can also be obtained
from a Taylor series using higher derivatives of $\vec{a}$ at $t=t_0$:
\begin{eqnarray}
 \vec{a}_1 &=& {1\over 6}(t-t_0)^3 \vec{a}_0^{(3)} +
              {1\over 2}(t-t_0)^2 \vec{a}_0^{(2)} + 
              (t-t_0)\vec{{\dot a}}_0 + \vec{a_0} \ ,\\
 \vec{{\dot a}}_1  &=& {1\over 2}(t-t_0)^2 \vec{a}_0^{(3)} +
               (t-t_0) \vec{a}_0^{(2)} + \vec{{\dot a}}_0 \ .
\label{4.1.3} 
\end{eqnarray}
If $\vec{a}_1$ and $\vec{{\dot a}}_1$ is known from direct summation
(from Eq. \ref{4.1.1}
using the predicted positions and velocities) one can invert the equations
above to determine the unknown higher order derivatives of
the acceleration at $t=t_0$ for the test particle:
\begin{eqnarray}
 {1\over 2} \vec{a}^{(2)} &=& -3 {\vec{a}_0 - \vec{a}_1 \over (t-t_0)^2}
   - {2\vec{{\dot a}}_0 + \vec{{\dot a}}_1 \over (t-t_0)} \\
 {1\over 6} \vec{a}^{(3)} &=& 2 {\vec{a}_0 - \vec{a}_1 \over (t-t_0)^3}
   - {\vec{{\dot a}}_0 + \vec{{\dot a}}_1 \over (t-t_0)^2} \ ,
\label{4.1.4} 
\end{eqnarray}
This is the Hermite interpolation, which finally allows to correct
positions and velocities at $t_1$ to high order from
\begin{eqnarray}
 \vec{x}(t) = \vec{x}_p(t) + {1\over 24}(t-t_0)^4 \vec{a}_0^{(2)}
                                 +{1\over 120}(t-t_0)^5 \vec{a}^{(3)} \ ,\\
 \vec{v}(t) = \vec{v}_p(t) + {1\over 6}(t-t_0)^3 \vec{a}_0^{(2)}
                                 +{1\over 24}(t-t_0)^4 \vec{a}_0^{(3)} \ .
\label{4.1.5} 
\end{eqnarray}

\begin{figure}
  \vspace{8truecm}
\includegraphics{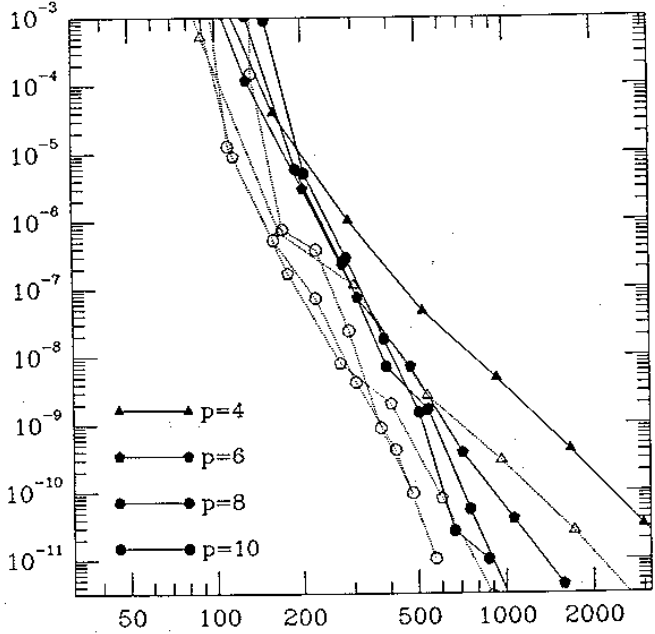}
 \caption{The relative energy error as the function of the number of steps.
 A time-step criterion using differences between predicted and corrected
 values is used, different from Eq. \ref{4.1.6}. Dotted curves are for
 Hermite schemes, solid curves for Aarseth schemes. The stepnumber $p$
 denotes the order of the integrator. From \cite{Makino91a}.
 }
  \label{Fig3Mak91}
\end{figure}

Taking the time derivative of Eq. \ref{4.1.5} it turns out that the
error in the force calculation for this method is $\co (\Delta t^4)$,
as opposed to the widely used leap-frog schemes, which have a force
error of $\co (\Delta t^2)$. Additional errors induced by approximate
potential calculations (particle mesh or \tree ) create potentially
even larger errors than that. In Fig. \ref{Fig3Mak91}, however, 
it is shown
that the above Hermite method used for a real $N$-body integration
sustains an error of $\co(\Delta t^4)$ for the entire calculation.
Many persons in the
world know as Aarseth scheme (in particular the
code version NBODY5 \cite{Aarseth85}) 
an integrator of the same order as the
Hermite scheme, but using only accelerations on four time points
instead of $\vec{a}$ and $\vec{\dot a}$ on two time points. As is
shown in \cite{Makino91a}, the Aarseth scheme is
$\co(\Delta t^4)$ as well,
but for the same number of time steps the absolute value of the
energy error (not its slope) is clearly smaller in the Hermite scheme.
This means that for a given energy error the Hermite scheme allows
timesteps which are larger by some factor of order unity depending
on the parameters of the system under study. The
Hermite scheme has been commonly adopted during the past years, because
it needs less memory, and allows slightly larger timesteps. More
importantly, after the addition
of a hierarchical (as opposed to individual) time step scheme it is
well suited for parallelization on modern special and general purpose
high performance computers \cite{SpurzemB98}. The timestep scheme will
be discussed now.

\subsection{Choice of Timesteps -- Parallelization}

\cite{Aarseth85} provides an empirical timestep criterion
\begin{equation}
\Delta t = \sqrt{\eta { \vert\vec{a}\vert \vert\vec{a}^{(2)}\vert
                          + \vert\vec{{\dot a}}\vert^2 \over
                          \vert\vec{{\dot a}}\vert \vert\vec{a}^{(3)}\vert
                          + \vert\vec{a}^{(2)}\vert^2 }} \ . 
\label{4.1.6} 
\end{equation}
The error is governed by the choice
of $\eta$, which in most practical applications is taken to be
$\eta = 0.01 - 0.04$.
It is instructive to compare this with the inverse square of the
curvature $\kappa$ of the curve $\vec{a}(t)$ in coordinate space
\begin{equation}
{1\over\kappa^2} = 
 {1+\vert\vec{\dot a}\vert^2\over\vert\vec{a}^{(2)}\vert^2} \ .
\end{equation}
Clearly under certain conditions the time step choice Eq. \ref{4.1.6}
becomes similar to choosing the timestep according to the curvature of
the acceleration curve; since it was determined just empirically, however,
it cannot generally be related
to the curvature expression above. In \cite{Makino91a} a different time
step criterion has been suggested, which appears simpler and more
straightforwardly defined, and couples the timestep to the difference
between predicted and corrected coordinates. 
The standard Aarseth time step criterion Eq. \ref{4.1.6} has been used in 
most $N$-body simulations so far
(but compare the discussion in \cite{Sweatman94}).

\begin{figure}
  \vspace{7truecm}
\includegraphics{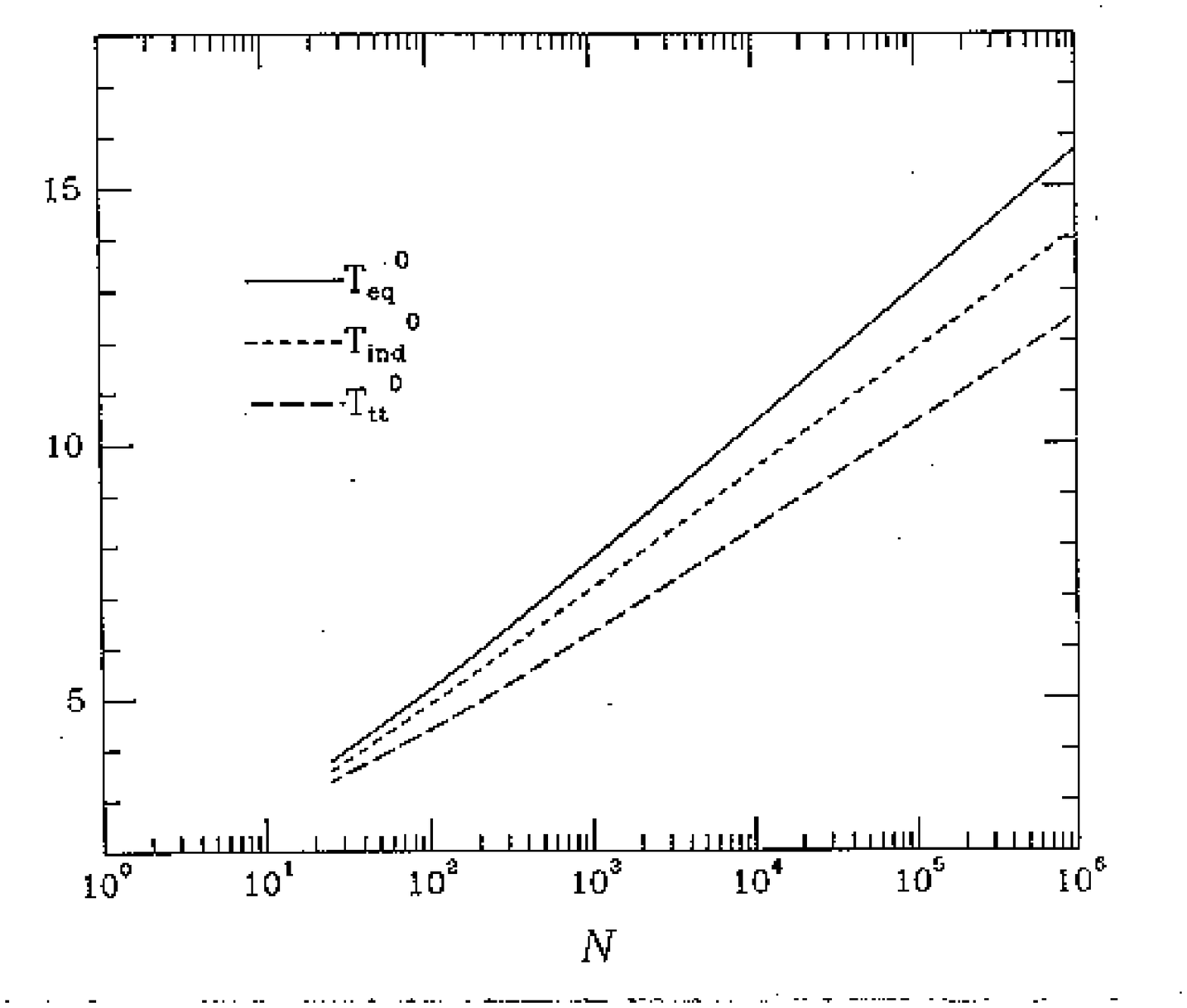}
\caption{The logarithm of $T$, the total computing time required to advance
the $N$-body system for one crossing time plotted as a function of the
particle number $N$ for the equal time step scheme $T_{\rm eq}$, the
individual time step scheme $T_{\rm ind}$ and the Ahmad-Cohen
neighour scheme with two levels of individual time steps $T_{\rm tt}$.
The unit of computing time is the time required to calculate the
force between a pair of particles. The system is assumed to be homogeneous.
From \cite{MakinoH88}.}
\label{Fig1MH88}
\end{figure}

\begin{figure}
  \vspace{7truecm}
\includegraphics{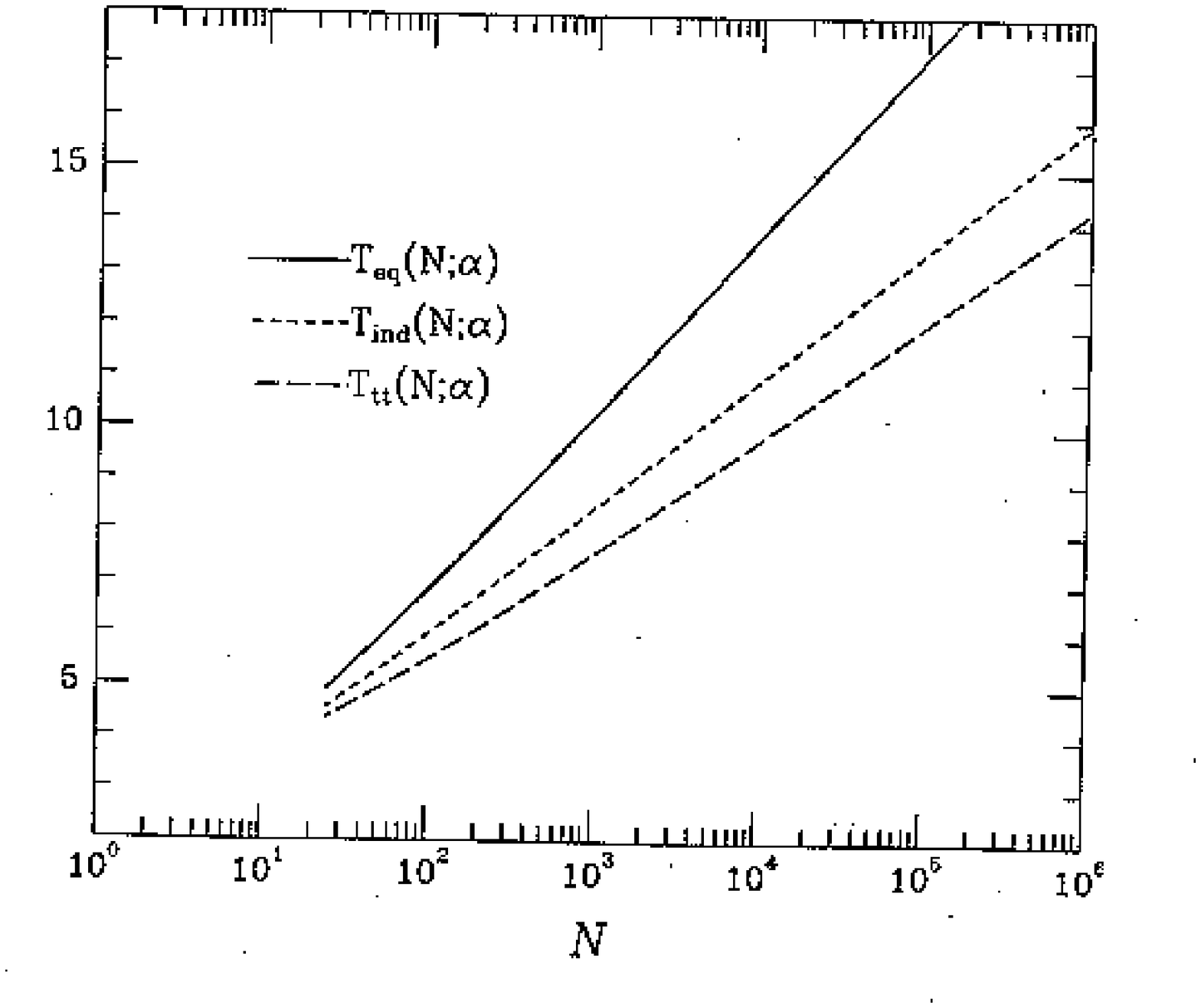}
\caption{As Fig. \ref{Fig1MH88}, but for the system with a power-law
density distribution $\rho\propto r^{-2.25}$. From \cite{MakinoH88}.}
\label{Fig2MH88}
\end{figure}

Since the position of all field particles can be
determined at any time by the low-order prediction Eq. \ref{4.1.2}, the
time step of each particle (which determines the time at which the
corrector Eq. \ref{4.1.5} is applied) can be freely chosen according to
the local requirements of the test particle; the additional error induced
due to the use of only predicted data for the full $N$ sums of Eq. \ref{4.1.1}
is negligibly small, for the benefit of not being forced to keep all
particles in lockstep. Such an individual time step scheme is in particular
for non-homogeneous systems very advantageous, as was quantitatively
pointed out by \cite{MakinoH88}.
Particles in the high density core of a star clusters need to be updated
much more often than particles on orbits very far from the centre. 
They show that the gain
in computational speed due to the individual time step scheme (as compared
to a lockstep scheme where all particles share the minimum required time step)
is of the order $N^{1/3}$ for homogeneous and $N^1$ for strongly spatially
structured systems; we show their results as Figs. \ref{Fig1MH88}, 
\ref{Fig2MH88}. 


For the purpose of vectorization and parallelization it is better not
to have the particles continuously distributed on a time axis. Consequently,
\cite{Makino91b} uses a hierarchical scheme, still on the basis of
Eq. \ref{4.1.6}; but a change of the timestep is considered only
if that equation yields a variation of $\Delta t$ compared
to the last step by more than a factor of 2 (increase or decrease).
If this is the case a variation by $2$ is applied only. Thus
in model units all timesteps are selected from the set 
$\{2^{-i}\vert i=0,...i_{\rm max}\}$ with $k = i_{\rm max}$ determined by the
condition that $\Delta t_{\rm min} > 2^{-i_{\rm max}}$ for
the minimum timestep $\Delta t_{\rm min}$ determined from Eq. \ref{4.1.6}.
For core collapse simulations of star clusters of a few ten thousand 
particles $i_{\rm max}$ goes up to about 20; empirically and
theoretically \cite{MakinoH88} $\Delta t_{\rm min}\propto N^{-1/3}$,
so for large $N$ $i_{\rm max}$ becomes larger, however, on
the other hand, how large $i_{\rm max}$ grows for fixed $N$
depends on the selected criteria for so--called KS regularisation of
perturbed two--body motion (see below). The implementation of the
block step scheme indeed uses an even stronger condition than the above
described one, it is demanded that not only the time steps, but also
the individual accumulated times of each particles are commensurate
with the timestep itself.
This ensures that for any particle $i$ and any time $T_i = t_i + \delta t_i $
{\it all} particles with $\delta t_j < \delta t_i $ have
for their own time $T_j = t_j + \delta t_j = T_i $, where the last
equality is the non--trivial one. Such procedure is important for
the parallelization of the algorithm. For example it has as a consequence
that at the big time steps always huge groups of particles are due
for correction, sometimes even all particles (at the largest steps).
Such scheme
allows an efficient parallelization of all operations necessary
for calculation of $\vec{a}$ and $\vec{\dot a}$ and for the update of
particle positions and velocities (corrections). Special purpose
computers have been built tailored to the Hermite codes, which are
denoted as {\sc HARP} (``Hermite Accelerator Pipeline'') boards and stem
from the bigger {\sc GRAPE}--family \cite{Sugimotoetal90,Makinoetal97}. Such
{\sc HARP}--boards have been made available also at some places outside Japan,
including ``Astronomisches Rechen--Institut'' Heidelberg (for an
application see e.g. \cite{TheisSp99}).

Another refinement of the Hermite or Aarseth ``brute force'' method is
the two-time step scheme, denoted as neighbour or Ahmad-Cohen scheme
\cite{AhmadC73}. For each particle a neighbour radius is defined,
and $\vec{a}$ and $\vec{\dot a}$ are computed due to neighbours and
non-neighbours separately. Similar to the Hermite scheme the
higher derivatives are computed separately for the neighbour
force (irregular force) and non-neighbour force (regular force). 
Computing two timesteps, an irregular small $\Delta t_{\rm irr}$ and 
a regular large $\Delta t_{\rm reg}$, from
these two force components by Eq. \ref{4.1.6} yields a timestep
ratio of $\gamma := \Delta t_{\rm reg}/\Delta t_{\rm irr}$ being in
a typical range of 5--20 for $N$ of the order $10^3$ to $10^4$. The
reason is that the regular force has much less fluctuations than
the irregular force. 
The Ahmad-Cohen neighbour scheme is implemented in a self-regulated way,
where at each regular time-step a new neighbour list is determined using a given
neighbour radius $r_{si}$ for each particle. If the neighbour number found
is larger than the prescribed optimal neighbour number, the neighbour radius
is increased or vice versa. In 
\cite{Aarseth85,MakinoH88} more complicated algorithms to adjust
the neighbour radius are described. On the
contrary to \cite{MakinoH88}, who find an optimal
neighbour number of $N_{n,\rm opt} \propto N^{3/4}$
we find that adopting a constant neighbour number of the
order of $20-50$ is sufficient at least up to $N=50000$. The reason is that
by using special purpose machines or parallelization for parts of
the code, an optimal neighbour number is not well defined, so the
neighbour number can be selected according to accuracy and efficiency
requirements \cite{SpurzemB98}. After each regular time step the new
neighbour list is communicated along with the new particle positions
to all processors of the parallel machine, thus making it possible to
do the irregular time step in parallel as well.

\begin{figure}
  \vspace{7truecm}
\includegraphics{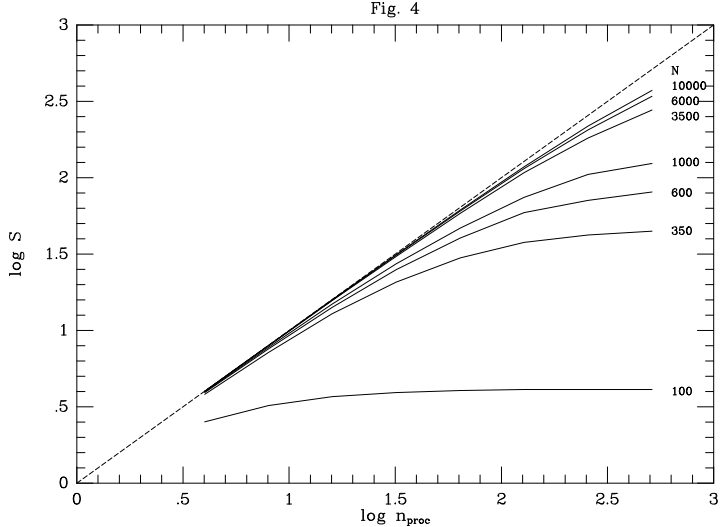}
 \caption{Theoretical speedup (neglecting
 communication) of regular force calculation as a function
 of processor number for varying particle number $N$.
 The dashed line is the ideal maximum speed up which could
 be reached on a given processor number.}
  \label{Fig1}
\end{figure}

\begin{figure}
  \vspace{7truecm}
\includegraphics{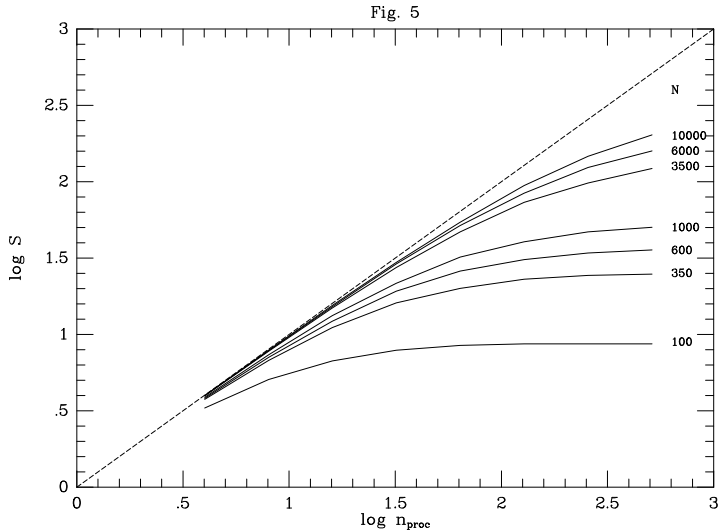}
 \caption{As Fig. \ref{Fig1}, for the irregular (neigbour) force
 calculation.}
  \label{Fig2}
\end{figure}

\begin{figure}
  \vspace{7truecm}
\includegraphics{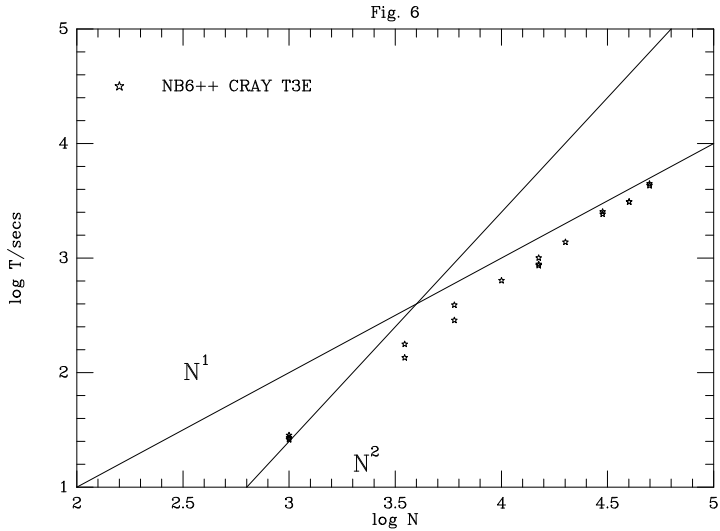}
\caption{
CPU time needed for one $N$-body time unit as
 a function of particle number $N$ using NBODY6++ on
 the CRAY T3E.
 The collection of data points includes
 runs with varying average neighbour number and processor/pipeline number,
 starting from 8 for low $N$ up to 512 for the largest $N$,
 which are not individually discriminated in the figure.}
\label{Fig3}
\end{figure}

Using a two-time step or neighbour scheme
again increases the computational
speed of the entire integration by a factor of at least proportional
to $N^{1/4}$ \cite{Makino91a}. Both the regular and
irregular timesteps are arranged in the
hierarchical, commensurable way, and the total inherent parallelism
in the resulting algorithm is depicted in Figs. \ref{Fig1}, \ref{Fig2} from
\cite{SpurzemB98} for the irregular and the regular step. One can
see that even for moderate particle numbers of $10^4$ particles
some 512 processors could be used efficiently.
Sometimes there are only very few particles in the smallest steps to be
integrated, which one might consider as 
being very prohibitive for parallelization.
However, due to the large number of medium and large size blocks this
effect is negligible for the overall performance. It causes however,
the saturation in the curves in Figs. \ref{Fig1} and
\ref{Fig2} which defines the
limit for the number of processors useful for a given particle number $N$.
By using more and more processors in the parallel execution one
finds that the asymptotic scaling of the ``brute force'' $N$-body problem
can be reduced effectively to an $N$ scaling (Fig. \ref{Fig3}). But in our
present implementation the parallelization is done only according to
parallel sections (do loops) in the code; there is no domain decomposition
(distributing particles on the processor). Thus at the end of any
timesteps new results have to be broadcast to all other processing units.
A systolic algorithm is used for that which scales linearly in communication
time with the number of processors. It is interesting to note an
approach suggested by molecular dynamicists to use a new kind of
hyper-systolic communication algorithm, which scales only by the
square root of the processor number \cite{Lippertetal96,Lippertetal98}.
Presently we think that hyper-systolic algorithms can efficiently be used only
if the sum over all particles for the acceleration and its time
derivative (Eq. \ref{4.1.1}) should be directly parallelized. The number
of interprocessor communications $N_{\rm comm}$ for the hyper-systolic
algorithm is of the order $N\sqrt{n_{\rm PE}}$;
on the other hand our algorithm, which we would
like to call here ``parallel group execution algorithm'' \cite{SpurzemB98}, 
has a
scaling $N_{\rm comm}\propto N^{2/3}n_{\rm PE}$, 
because only subgroups of particles,
whose size scales with $N^{2/3}$ have to be communicated across the
processor network. In other words, asymptotically (above
some critical particle number as a function of $n_{\rm PE}$, the hyper-systolic
algorithm should lose against the parallel group execution algorithm.
However, these questions have not yet been examined in detail, for example
what the critical $N$ really are and which algorithm is more efficient
for practically useful particle numbers of today. This is subject of
present and future work.

If the two-body force between any pair of particles becomes dominant
their (perturbed) relative motion is integrated in special regularized 
coordinates
(taking into account perturbations from field particles), in which
the singularity of the two-body motion is transformed into a slowly
varying parameter (the binding energy) and does not occur in the
integration variables. The rest of the $N$-body simulation generally
regards the regularized pair as a compound particle located at the
position and moving with the velocity of its centre of mass, except
in the case when a perturber moves very close to a regularized pair (in
such cases the pair is resolved). It was already discovered in the
earliest published $N$-body simulations that the formation of close and
eccentric binaries
occurs as the rule rather than as an exception and that it was particularly
difficult to accurately integrate them \cite{vHoerner60,vHoerner63}.
As a consequence two-body, three-body and chain regularizations were
developed and implemented in order to accurately and efficiently integrate
star clusters including all their close binaries, triples and hierarchical
subsystems. An excellent account of regularization, historically and
scientifically, can be found in \cite{Mikkola97b}. Most recent developments
are the slow-down treatment of tight binaries \cite{MikkolaA96} and a new
method to gain accuracy and exact solutions in the unperturbed case using
Stumpff functions \cite{MikkolaA98}.

Recently the necessity of regularization was challenged and its replacement
by a binary tree structure for hierarchical systems with relative coordinates
has been suggested \cite{MakinoT98}. However, the regularisation procedure
is undisputedly much more efficient and accurate for highly eccentric
binaries, and the new method has not yet been widely applied and proven
to work through the most delicate phase of core bounce and post-collapse
evolution in point-mass systems or systems with many primordial hard 
binaries.

\subsection{Exponential Instability, Validity of $N$-body simulations,
Planetary System Integrations}

Concerning the algorithms explained in the previous paragraph the
direct $N$-body simulation may turn out to be the most reliable (although
computationally most expensive) way to simulate the dynamical
evolution of a gravitating system consisting of $N$ point masses. 
It does not involve any serious approximations and assumptions, 
as e.g. the Fokker-Planck approximation in the
gaseous models. By reducing the $\eta$-values any accuracy can
be achieved in principle, 
as far as the globally conserved quantities (energy,
angular momentum) are concerned. However, for a system with $N$
particles phase space has $6N$ dimensions, and a check of say
energy and angular momentum alone only checks whether the numerically
calculated system remains within the allowed $6N-4$ dimensional hypervolume.
There is no a priori information how ``exact'' the individual trajectories
are reproduced in the simulation.
\cite{Miller64} pointed 
out that, due to repeated close encounters occurring between
particles initial configurations that are very
close to each other, quickly diverge in their evolution from each other.
He could show that the separation in phase space of two trajectories
increases exponentially with time, or with other
words, the evolution of the configuration is extremely sensitive
to initial conditions (particle positions and velocities). The timescale
of exponential instability is as short as a fraction of a crossing time,
and the accurate integration of a system to core collapse would require
of order $\co(N)$ decimal places \cite{GoodmanHH93,KandrupMS94}.
Those papers
argue that the problem is caused by two-body encounters, but chaotic
orbits in non-integrable potentials can be a source of exponential
instability and thus cause unreliable numerical integrations as well.

However, the situation is not as bad as it seems. $N$-body simulations for
star clusters or galactic nuclei
do not really exploit the detailed configuration space of all particles.
Quantities of interest are global or somehow averaged quantities,
like Lagrangian radii or velocity dispersions averaged in certain
volumes. As it was nicely demonstrated in the pioneering series
of papers by \cite{GH1,GH2,GH3,GH4} such results are
not sensitive to small variations of initial parameters. They
took statistically independent initial models (positions and velocities
at the beginning selected by different random number sets) and showed
that the ensemble average of the dynamical evolution of the system
always evolved predictably and in remarkable accord with results
obtained from the Fokker-Planck approximation. The method was also
partly and successfully
used in \cite{SpD}, which focused on the evolution of anisotropy 
and comparisons with the anisotropic gaseous models of the author of
this paper. 

As a consequence, it should be remembered, however, that great
care has to be taken when interpreting results of $N$-body
simulations on a particle by particle basis, for example
determining rates of specific types of encounters, which could
produce mergers in a large direct $N$-body model. 

The long-term behaviour of dynamical systems as the solar system are
being studied by $N$-body simulations as well, but clearly there are
much higher requirements on the accuracy of the individual orbits in contrast
to the star cluster problem. Therefore for the solar system dynamics
symplectic methods, using a generalized leap-frog, like the widely used
Wisdom-Holman symplectic mapping method \cite{WisdomH91} are the
standard integration method. As a non-exhaustive
reference the reader might look into a recent study of the relation
between the earth-moon system and the stability of the inner solar system
using this method \cite{Innanenetal98} and a contemporary review
\cite{DuncanQ93}. Symplectic mapping methods
do not show secular errors in energy and angular momentum. However, in their
standard implementation they require a constant timestep. A generalization
using a time transformation simultaneously with the generalized leap-frog
has been suggested which can cope with variable timesteps \cite{Mikkola97a}.
Another more practical approach to strongly reduce secular errors
is to enforce a time-symmetric scheme by making the timesteps reversible
through an iteration \cite{Hutetal95,Funatoetal96}. How well this generally
works and its relation to symplectic schemes is presently not clear. 
In \cite{MikkolaA98} it is stressed that even with a newly applied
classical method secular errors in the integration of close binaries
can be strongly reduced. One should keep in mind though, that the
$N$-body integration schemes discussed in this paper
yield excellent results in the star 
cluster research (see Sect. 4) but are unsuitable for long-term solar
system studies, because they generally have secular errors, although small.
As outlined above in star cluster simulations
the secular errors are being kept small relative
to typical values of energy and angular momentum and
an accurate reproduction of all individual stellar orbits is not 
generally required.

\subsection{What about \tree- and fast multipole codes?}

\begin{figure}
\vspace{7truecm}
\includegraphics{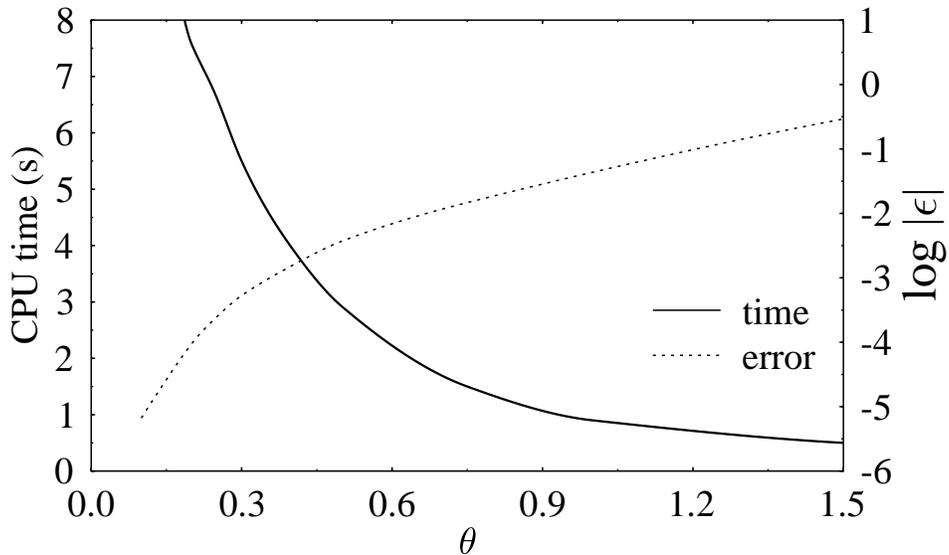}
\caption{Tradeoff between CPU time per step and average force error for
the \tree-code with monopole terms only. 
From Fig. 4.11. of \cite{PfalznerG96}.}
\label{FigPG411}
\end{figure}

\begin{figure}
\vspace{7truecm}
\includegraphics{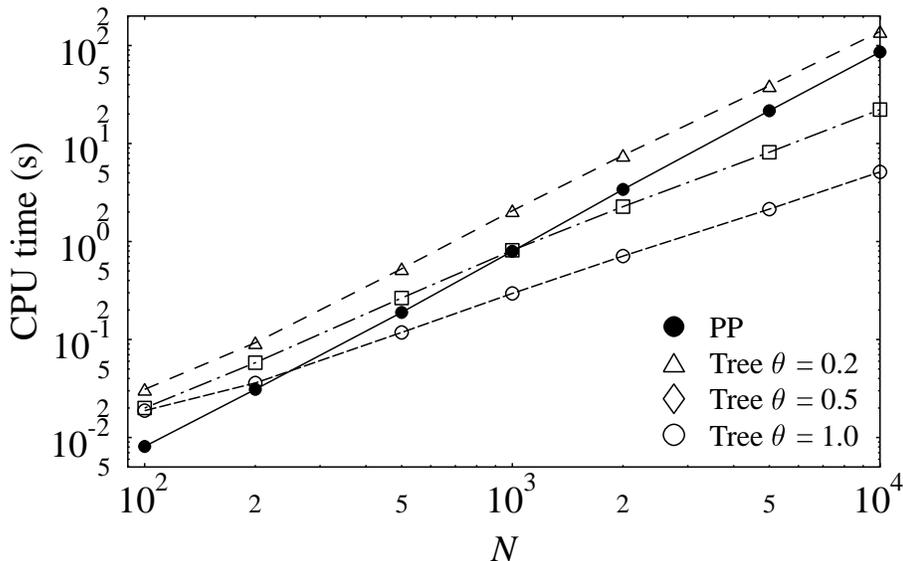}
\caption{CPU time per step versus particle number $N$ for \tree-codes with
varying opening parameter $\theta$ and a direct full ``brute-force'' labelled
with PP in the figure (for ``particle--particle''). From Fig. 4.9. of
\cite{PfalznerG96}.}
\label{FigPG49}
\end{figure}

Finally, remarks shall be made on two very widely used algorithms
to compute gravitational potentials from particle distributions
namely the \tree- and fast multipole (FMP) algorithms. The \tree - method
of \cite{BarnesH86} divides the system into hierarchical cells. The
mutual interaction between particles or cells is resolved only if
the opening parameter $\theta = r/d$, where $r$ is the distance to
and $d$ a size scale of the cell under consideration, is smaller than
a prescribed critical $\theta_{\rm crit}$. If the cell is not resolved
because $\theta < \theta_{\rm crit}$, there is still the option to
evaluate multipole moments of its internal mass distribution for the
interaction with external particles. As one can see from Fig. \ref{FigPG411},
a global accuracy requirement of 
$\Delta E/E \approx 10^{-5}$ demands $\theta_{\rm crit}\approx 0.2$,
a value much smaller than the usually efficient choice of 0.5--0.7,
at which the computational time scales approximately as $\co(N\ln N)$.
Looking then at Fig. \ref{FigPG49}, the computational
time for the \tree-code with $\theta_{\rm crit}\approx 0.2$
scales nearly as $\co(N^2)$, i.e. like a
``brute force'' algorithm. So for each particle number and required
accuracy one should carefully check whether a \tree-code or a direct
$N$-body code are the best choice.

\begin{figure}
\vspace{6truecm}
\includegraphics{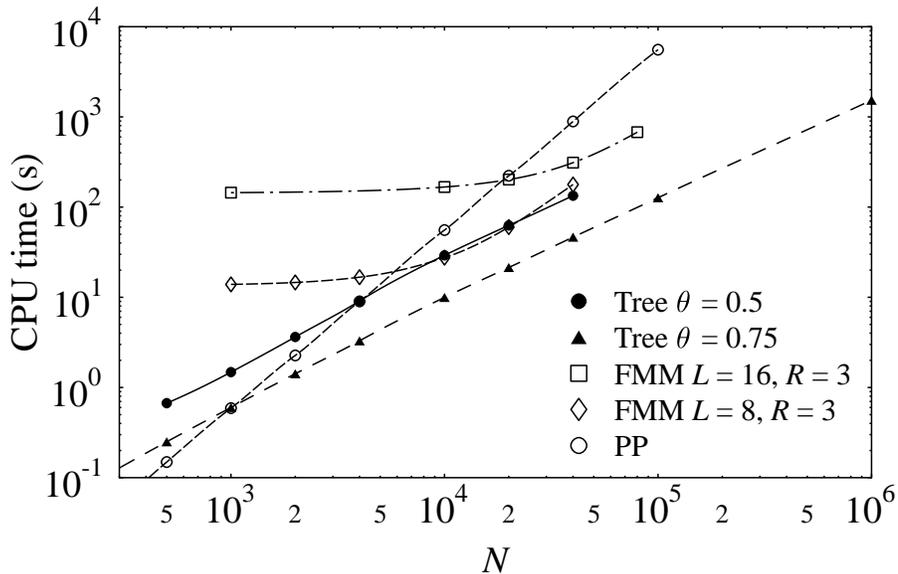}
\caption{Comparison of computational time as a function of particle number
$N$ between particle-particle, hierarchical \tree, and the fast multipole
code. From Fig. 7.14 of
\cite{PfalznerG96}.}
\label{FigPG714}
\end{figure}

Another \tree-based algorithm is the fast-multipole method (FMT) proposed
by \cite{Greengard87,GreengardR87}. The pair-wise
potential in Eq. \ref{4.1} is approximated by a multipole series, which
can be done for arbitrary precision if enough terms are included. The multipole
terms
used for different test particles can be transformed into each other by using
clever addition theorems for spherical harmonics, so the entire algorithm
scales in its computational demand with $\co(N)$ only. Higher precision
only changes the proportionality factor, not the scaling (as in the
case of the \tree-code, which effectively becomes a ``brute force'' code
if high enough accuracy is demanded. However, such a code is fine only for
homogeneous or nearly homogeneous systems, as they occur in plasma
physics. In all cases where there
is strong spatial structure, like in astrophysical star clusters, 
\cite{MakinoH88} have demonstrated
that the use of an individual time step scheme in an $\co(N^2)$
code gains a factor at least $\propto N$ in efficiency. So, asymptotically a
``brute-force''
integrator with individual timesteps is more efficient than an FMT integrator.
The latter is based on an equal timestep for all particles (otherwise it
would lose its $\co(N)$ property; so both codes have asymptotically
the same $N$ scaling, but then the overhead (proportionality) factor is
much smaller in the direct force summation than in the multipole evaluation.
This can be seen also from Fig. \ref{FigPG714} in \cite{PfalznerG96} for 
low $N$.
For the direct calculation method in this plot, the individual time step
scheme is not taken into account.

The information contained in the previous paragraphs, complemented by some
additional details and references, 
which will not be elaborated in more detail here,
are presented in an overview in Table \ref{Tab1}. 
It is divided into three boxes,
the first for the mesh or series evaluation codes, which do not contain
particle-particle forces and thus are not appropriate for direct modelling
of relaxing systems. The second box contains the classical direct
``brute force'' $N$-body codes, whereas the third one contains algorithms
which cannot clearly be counted to one of the other two groups.

\begin{center}
\begin{table}
\caption{
\centerline{Algorithms for $N$-body Simulations}
\centerline{$N$: particle, $N_n$: characteristic neighbour number}
\centerline{$n_{\rm c}$: number of grid cells in one dimension,
$nlm$: order in 3D series evaluation}}
\label{Tab1}
\begin{tabular}{|cccc|}
\hline
Acronym & Algorithm & Scaling & Comments \\
\hline
PM      & Particle Mesh      & $N$ $n_{\rm c}^3\log_2 n_{\rm c}^3$ ${}^{(1)}$ &
                                                    fixed geometry  \\
FMP     & Fast Multipole     & $N$  $nlm$ & req. equal $\Delta t$ \\
SCF     & Self-Consistent Field &$N$  $nlm$ & 
                                              series evaluation ${}^{(2)}$\\
\hline
{\sc Nbody1}   & Aarseth     &  $N^2$   & ITS, softening \\
{\sc Nbody1++} & Hermite     &  $N^2$   & HTS, softening \\
{\sc Nbody2}   & Aarseth, AC & $NN_n+N^2/\gamma $ & ITS, softening, ${}^{(3)}$\\
{\sc Nbody3}   & Aarseth     &  $N^2$   & ITS, KS-reg. \\
{\sc Nbody4}   & Hermite     &  $N^2$   & HTS, KS-reg.  \\
{\sc Nbody5}   & Aarseth, AC & $NN_n+N^2/\gamma $ & ITS, KS-reg., ${}^{(3)}$\\
{\sc Nbody6}   & Hermite, AC & $NN_n+N^2/\gamma $ & HTS, KS-reg., ${}^{(3)}$\\
{\sc Nbody6++} & parallel {\sc Nbody6}  & $NN_n+N^2/\gamma $ &
 HTS, KS-reg., ${}^{(3,4)}$ \\
{\sc Kira}     & Hermite     &  $N^2$   &  HTS, ${}^{(5)}$\\
\hline
{\sc Tree}     & \tree-code  & $N\ln N$ & $N^2$ for high accuracy \\
${\rm P}^3{\rm M}$ & Part.-Part. PM & $N_n^2$ 
$n_{\rm c}^3\log_2 n_{\rm c}^3$ ${}^{(1)}$ &
                                            fixed geometry ${}^{(6)}$ \\
\hline

\end{tabular}

softening: singularity in pairwise potential removed by softening parameter
$\varepsilon$

ITS: Individual Time Step Scheme

HTS: Hierarchical Block Time Step Scheme

KS-reg.: KS regularization of perturbed two- and hierarchical $N$-body motion
 \cite{KustaanheimoS65,MikkolaA98}

AC: Ahmad-Cohen neighbour scheme \cite{AhmadC73}

${}^{(1)}$ Discrete FFT on regular 3D mesh with $n$ linear
mesh points assumed

${}^{(2)}$ Sufficient Accuracy requires appropriate basis function set 
           \cite{HernquistO92} 

${}^{(3)}$ $\gamma$: ratio of regular to irregular time step

${}^{(4)}$ speedup by parallel execution not contained in scaling, see
  \cite{SpurzemB98}

${}^{(5)}$ New high accuracy Hermite code based on {\sc Starlab} 
 \cite{McMillanH96,Portegetal98}

${}^{(6)}$ with hierarchically nested adaptive grids used for cosmological 
 simulations \cite{PearceC97}

\end{table}
\end{center}

\section{Application to Star Clusters}

Since this article is focused on the physical and numerical methods
of calculating the evolution of
relaxing star clusters, only a brief
account of some of the physical problems and challenges will be given
here, which have been and will be tackled by the previously described
models. Despite of a wealth of beautiful observational data provided
by e.g. Hubble space telescope observations of globular clusters some
of the fundamental questions related to the validity of the $N$-body 
approach and the other approximate models still deserve attention as they
can lead to very fascinating general questions regarding the
thermodynamical behaviour of large $N$-body systems.

\begin{figure}
  \vspace{7truecm}
\includegraphics{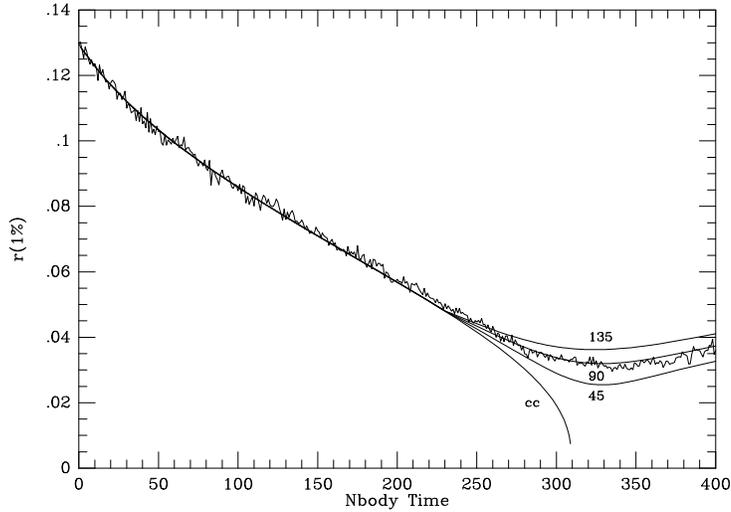}
 \caption{Evolution of the 1\% Lagrangian radius in the averaged
 $N=1000$
 $N$-body model in comparison to the anisotropic gaseous model
 for different strength of the binary energy generation parameter
 $C_b$ (see Eq. \ref{2.3.16}).
 The subscript $cc$
 indicates a pure core collapse gaseous model without binary heating.}
  \label{Fig4}
\end{figure}

\begin{figure}
  \vspace{7truecm}
\includegraphics{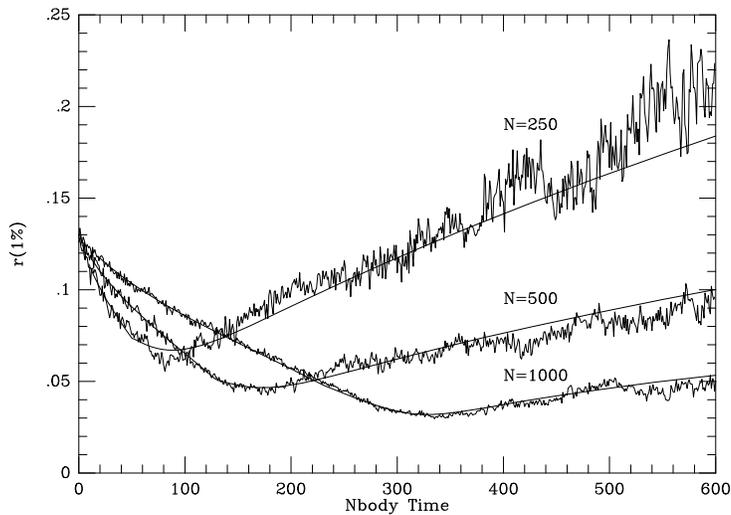}
\caption{Evolution of the 1\% Lagrangian radius in the
 averaged 1000,
 50, 250 body models
 in comparison to the anisotropic gaseous models using
 $C_b = 90$, 70, 55, respectively.}
  \label{Fig5}
\end{figure}

A series of papers has been devoted to the comparison of ensemble
averaged $N$-body simulations ($N\le 2000$) with the expectations
derived from Fokker-Planck or gaseous models \cite{SpD,GH1,GH2,GH3,GH4}.
Here we show as an example, in Figs. \ref{Fig4} and \ref{Fig5}, the excellent
agreement reached between the anisotropic gaseous model and the
ensemble averaged $N$-body system. The models started
with an initial Plummer model and follow the
core collapse induced by heat conduction and the
post-collapse evolution due to formation and hardening
of three-body hard binaries. The agreement of both types of
models mutually supports both sides: it shows that by ensemble averaging,
the exponential instability of the $N$-body system does not spoil
the physically correct behaviour of the system. It also demonstrates that
the Fokker-Planck approximation, especially with its underlying assumption
of strict spherical symmetry and dominance of small-angle two-body
encounters for relaxation (i.e. neglectance of collective processes), 
is correct.
It also shows that the very simple algorithm to describe the heating
provided by the formation of close three-body binaries and their subsequent
hardening by superelastic binary-single star encounters, which was
first introduced into the gaseous models by \cite{BettwieserSu84}, provides
a surprisingly good description of the real processes in the 
average $N$-body system. The cited paper ignited a discussion
over many years whether gravothermal oscillations, being a thermodynamic 
consequence of heat conduction by two--body relaxation, will prevail
in a real $N$-body system with all its stochastic fluctuations.
The question was settled after an $N$-body simulation
on the massively parallel Teraflop GRAPE machine \cite{Makinoetal97}
using a high-accuracy Hermite scheme, as described above, became available.
Gravothermal
oscillations were found in a very large $N=32000$ particle simulation 
\cite{Makino96}.

\begin{figure}
  \vspace{7truecm}
\includegraphics{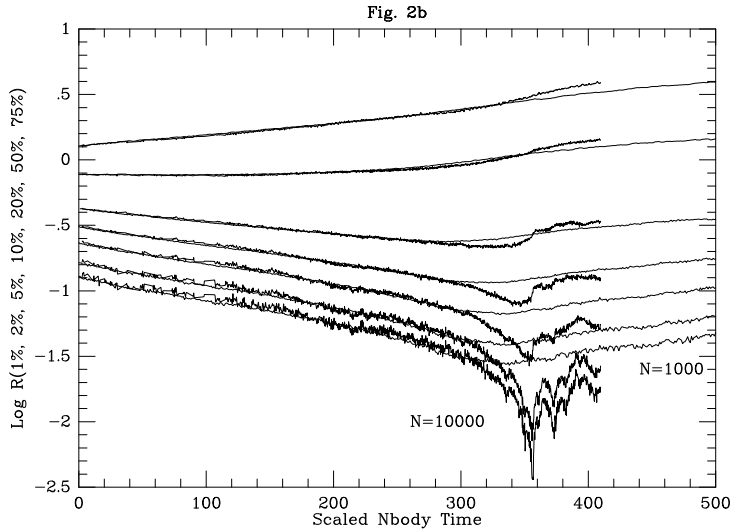}
\caption{
Lagrangian radii containing the indicated
fraction of total mass as a function of time for a single $10^4$-body
simulation (fluctuating curves) compared to an averaged $N=1000$ simulation of
\cite{GH1,GH2}. Times scaled as explained in main text.}
  \label{Fig6}
\end{figure}

In Fig. \ref{Fig6} we show a striking example of the validity of the
Fokker-Planck approximation even for a single large direct $N$-body-simulation,
here using {\sc Nbody5}, an Aarseth scheme (see Table \ref{Tab1}), for
10000 particles, a model simulation again starting with Plummer's model
and undergoing core collapse and core bounce due to hard binaries \cite{SpG}.
The average $N=1000$ particle model by \cite{GH1}
has been taken and its time was scaled with the factor $N/\ln(\gamma N)$,
which is the scaling of the standard two--body relaxation time Eq. \ref{1.4}.
An excellent match between the evolution of the Lagrangian radii for
the two systems occurs after such scaling, proving that it is indeed
the standard relaxation which dominates the pre-collapse evolution.
The differences between the two systems show up at the moment of the
formation of the first three-body binaries, after which one expects
the evolution not to scale as the relaxation time. In simple terms,
the larger $N$, the less important are three-body effects as compared
to the global potential; hence for large particle numbers the system
collapses to higher densities and three-body effects finally dominate
because they depend on the third power of the particle density as
compared to the $n^2$ dependence of two-body relaxation.

\begin{figure}
\vspace{7cm}
\includegraphics{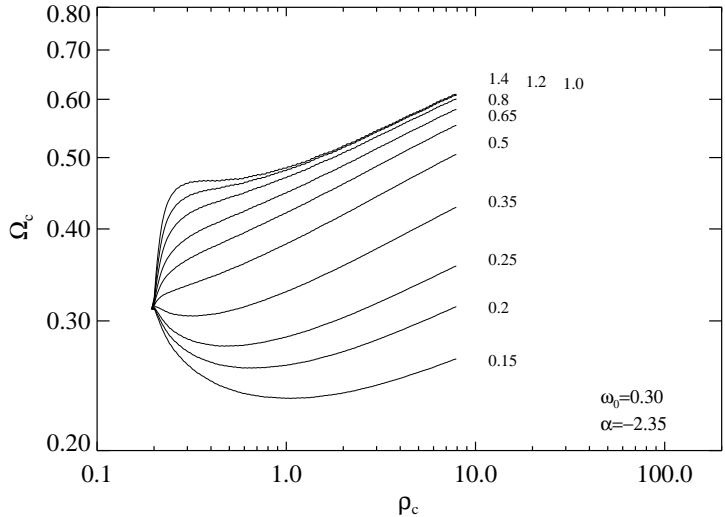}
\caption{ Evolution of central angular velocities versus central
density of all 10 mass bins
used in the calculation for an initial model with dimensionless
central potential $W_{0}=3$ and dimensionless angular velocity
$\omega_0=0.30$. The highest mass bin with $m=1.4~M_{\odot}$ is the
uppermost curve, while the lowest mass bin with $m=0.15~M_{\odot}$ is
the lowest curve. The parameters $W_0$ and $\omega_0$ refer
to an initial Michie-King model.
}
\label{Fig7}
\end{figure}

Finally, we show a result from \cite{Einsel98} in Fig. \ref{Fig7}, 
a new multi-mass model
using the orbit averaged Fokker-Planck approximation for axisymmetric
rotating relaxing star clusters. The standard effect of mass segregation
of the heavy masses is accompanied here by an acceleration of their
rotational speed as compared to the small masses. Such interesting dynamical
behaviour occurs just due to point-mass relaxation processes starting
with a very simple tidally truncated rotating King model without any mass or
rotational segregation. It is a yet unpublished generalization of
equal mass rotating star cluster models \cite{EinselSp99}. They neglect
the possible dynamical effect of non-classical third integrals, since
it is assumed that the distribution function depends on energy and
$z$-component of the angular momentum only. Such approximation needs
to be checked by direct $N$-body models, which is the subject of on-going work.
The results will also be important for the dynamical study of rotating
galactic nuclei containing massive star-accreting black holes.

The reader should be made aware of the
problem of scaling in the description of escaping stars from
globular clusters \cite{Portegetal98,AarsethH98} being tackled
by large direct $N$-body simulations and their comparisons with
approximate models. There are more challenges, like the inclusion
of many close binaries already originating from star formation processes
(for a calculation using {\sc Nbody5} see \cite{Kroupa95}, compare
also \cite{BonnellD98,Kroupaetal98} for the study of mass segregation in young
forming star clusters by means of direct $N$-body models).
Finite sizes of stars lead to merging in high-density phases and
cause population gradients and unusually high
frequencies of exotic objects like blue stragglers and pulsars in the
cores and haloes of globular clusters. Attempts to model all these
processes in direct $N$-body models, with as many ingredients and
realistic features included as possible are under way \cite{Aarseth96}.
Ultimately we will be able from such models to provide synthetic
observational data as e.g. color-magnitude diagrams. 

\section*{Acknowledgements}
Part of the work presented here was supported by DFG grants Sp 345/3-3
and 5-3. Computational time on the CRAY T3E parallel machines at
HLRZ J"ulich and HLRS Stuttgart is gratefully acknowledged. I thank
the ``Grapeyard'' in Tokyo (Jun Makino
and all other colleagues) for the
continuous support and cooperation in all aspects regarding the {\sc GRAPE}
hardware and its scientific use. I am very grateful to
Sverre Aarseth, for teaching me
the art of $N$-body simulations years ago and continuously supplying new
pieces of code and ideas. This work would not have been possible without
his support. Thanks go also to Sverre Aarseth, Hugh Couchman, and Pavel Kroupa
for helpful comments and suggestions.

%==============================================================================
% ==> entries to your list of references should appear between the
%      \begin{thebibliography} and \end{thebibliography} commands in the
%      order they appear in the text. Please use the \bibitem command
%      according to the following sample
%
%      \bibitem{Boyce89}
%      P.J. Boyce, R.J. Cohen, and W.R.F Dent,
%      An unusual molecular cloud complex in the Galactic Centre,
%      {\em Mon. Not. R. Astron. Soc. /\}, {\bf 239} (1989) 1013--1023
~
% ==> in the text you refer to this using
%      \cite{Boyce89}
%
%==============================================================================

\end{document}